\documentclass{emulateapj}
\usepackage{natbib}
\usepackage{amsmath}
\newcommand{\spider}{{\sc Spider}}
\newcommand{\boomerang}{{\sc Boomerang}}
\newcommand{\bicep}{{\sc Bicep}}
\newcommand{\biceptwo}{{\sc Bicep2}}
\newcommand{\WMAP}{{\sc WMAP}}
\newcommand{\Archeops}{{\sc Archeops}}
\newcommand{\synfast}{{\it synfast}}
\newcommand{\anafast}{{\it anafast}}
\newcommand{\healpix}{{\sc healpix}}
\newcommand{\Muller}{M\"{u}ller\ }
\newcommand{\fwhm}{\textsc{fwhm}}
\newcommand{\dust}{\mathrm{dust}}

\newcommand{\fds}{\mathrm{FDS}}
\newcommand{\thrD}{3\mathrm{D}}
\newcommand{\sky}{\mathrm{sky}}
\newcommand{\sech}{\mathrm{\,sech\,}}
\newcommand{\nn}{\nonumber}
\newcommand{\ud}{{\mathrm{d}}}
\newcommand{\vect}[1]{\ensuremath{\boldsymbol{#1}}}
\newcommand{\mat}[1]{\ensuremath{\textbf{\textsf{#1}}}}

\newcommand{\vx}{\vect{x}}
\newcommand{\mM}{\mat{M}}
\newcommand{\mR}{\mat{R}}

\newcommand{\beginfigure}{\begin{figure*}}
\newcommand{\efigure}{\end{figure*}}
\graphicspath{{./figs/pdf/}{./figs/eps/}}

\slugcomment{Submitted to ApJ}
\shorttitle{\spider\ Optimization II}
\shortauthors{O'Dea et al.}

\begin{document}
\title{\spider\ optimization II: optical, magnetic and foreground effects}

\author{
D.~T.~O'Dea,\altaffilmark{1}  
P.~A.~R.~Ade,\altaffilmark{2} 
M.~Amiri,\altaffilmark{3} 
S.~J.~Benton,\altaffilmark{4}
J.~J.~Bock,\altaffilmark{5,6} 
J.~R.~Bond,\altaffilmark{7} 
J.~A.~Bonetti,\altaffilmark{6} 
S.~Bryan,\altaffilmark{8} 
B.~Burger,\altaffilmark{3} 
H.~C.~Chiang,\altaffilmark{9} 
C.~N.~Clark,\altaffilmark{1}
C.~R.~Contaldi,\altaffilmark{1} 
B.~P.~Crill,\altaffilmark{5,6} 
G.~Davis, \altaffilmark{3}
O.~Dor\'e,\altaffilmark{5,6} 
M.~Farhang,\altaffilmark{7,10}
J.~P.~Filippini,\altaffilmark{5}  
L.~M.~Fissel,\altaffilmark{10} 
A.~A.~Fraisse,\altaffilmark{9}
N.~N.~Gandilo,\altaffilmark{10}
S.~Golwala,\altaffilmark{5} 
J.~E.~Gudmundsson,\altaffilmark{9} 
M.~Hasselfield,\altaffilmark{3} 
G.~Hilton,\altaffilmark{11} 
W.~Holmes,\altaffilmark{6} 
V.~V.~Hristov,\altaffilmark{5} 
K.~Irwin,\altaffilmark{11} 
W.~C.~Jones,\altaffilmark{9} 
C.~L.~Kuo,\altaffilmark{12} 
C.~J.~MacTavish,\altaffilmark{13} 
P.~V.~Mason,\altaffilmark{5} 
T.~E.~Montroy,\altaffilmark{8} 
T.~A.~Morford,\altaffilmark{5} 
C.~B.~Netterfield,\altaffilmark{4,10} 
A.~S.~Rahlin,\altaffilmark{9} 
C.~Reintsema,\altaffilmark{11} 
J.~E.~Ruhl,\altaffilmark{8} 
M.~C.~Runyan,\altaffilmark{5} 
M.~A.~Schenker,\altaffilmark{5} 
J.~A.~Shariff,\altaffilmark{10} 
J.~D.~Soler,\altaffilmark{10} 
A.~Trangsrud,\altaffilmark{5} 
C.~Tucker,\altaffilmark{2} 
R.~S.~Tucker,\altaffilmark{5} 
A.~D.~Turner,\altaffilmark{6} and
D.~Wiebe,\altaffilmark{3}
}
\altaffiltext{1}{Theoretical Physics, Blackett Laboratory, Imperial
  College, London, UK}
\altaffiltext{2}{School of Physics and Astronomy, Cardiff University, Cardiff, UK}
\altaffiltext{3}{Department of Physics and Astronomy, University of British
Columbia, Vancouver, BC, Canada}
\altaffiltext{4}{Department of Physics, University of Toronto, Toronto, ON,
Canada}
\altaffiltext{5}{Department of Physics, California Institute of Technology,
Pasadena, CA, USA}
\altaffiltext{6}{Jet Propulsion Laboratory, Pasadena, CA, USA}
\altaffiltext{7}{Canadian Institute for Theoretical Astrophysics, University
of Toronto, Toronto, ON, Canada}
\altaffiltext{8}{Department of Physics, Case Western Reserve University,
Cleveland, OH, USA}
\altaffiltext{9}{Department of Physics, Princeton University, Princeton, NJ, USA}
\altaffiltext{10}{Department of Astronomy and Astrophysics, University of Toronto, Toronto, ON,
Canada}
\altaffiltext{11}{National Institute of Standards and Technology, Boulder, CO, USA}
\altaffiltext{12}{Department of Physics, Stanford University, Stanford, CA, USA}
\altaffiltext{13}{Kavli Institute for Cosmology, University of Cambridge, Cambridge, UK }

\begin{abstract}
 \spider\ is a balloon-borne instrument designed to map the polarization of the cosmic
microwave background (CMB) with degree-scale resolution 
over a large
fraction of the sky.  \spider's main goal is to measure the amplitude
of primordial gravitational waves through their imprint on
the polarization of the CMB if the tensor-to-scalar ratio, $r$, is
greater than $0.03$.  To achieve this goal, instrumental systematic
errors must be controlled with unprecedented accuracy.  Here, we build
on previous work to use simulations of \spider\ observations to examine
the impact of several systematic effects that have been characterized
through testing and modeling of various instrument components.  In
particular, we investigate the impact of the non-ideal spectral
response of the half-wave plates, coupling between focal plane
components and the Earth's magnetic field, and beam mismatches and
asymmetries.  We also present a model of diffuse polarized foreground
emission based on a three-dimensional model of the Galactic magnetic
field and dust, and study the interaction of this foreground
emission with our observation strategy and instrumental effects.  We
find that the expected level of foreground and systematic
contamination is sufficiently low for \spider\ to achieve its science
goals.
\end{abstract}


\section{Introduction}
Observations of the cosmic microwave background (CMB) have been
central to the development of a standard cosmological model over the
past few decades.
A key component of this model is \textit{inflation}, a period of
accelerated expansion that occurred very early in the history of the
universe, which enables the model to reproduce the flatness and
isotropy observed today.
%
Importantly, inflation also predicts a nearly scale-invariant spectrum
of primordial density (scalar) perturbations, in excellent agreement
with recent observations of the anisotropies in the intensity of the
CMB \citep{Larson:2010gs}. 

Although the inflationary hypothesis was first put forward several
decades ago \citep{physRevD.23.347}, the details of the underlying
physics that drives the expansion are still uncertain. Furthermore,
despite the advances made through recent observations of the CMB, the
constraints on the general parameters used to describe inflation are
weak. As a result, a plethora of plausible inflationary scenarios have
been advanced, drawing on a wide range of proposed physical
mechanisms.


In addition to the density perturbations that seed large-scale
structure, inflation generates a stochastic background of
gravitational waves (tensor perturbations) that leave a unique imprint
on the polarization of the CMB by introducing a curl or ``$B$-mode''
component.  A detection of the $B$-mode signature would provide strong
evidence for inflation and a measure of its energy scale,
representing a major breakthrough in cosmology.  The combination of
this measurement with our current knowledge of the scalar
perturbations would begin to elucidate the dynamics of inflation and
place strong constraints on the underlying physics.  As the current
upper limit on the energy scale of inflation is around $10^{16}$\,GeV,
close to the GUT scale, measuring the $B$-mode signature of
gravitational waves in the CMB provides a unique opportunity to probe
physics at energies far beyond the reach of terrestrial high-energy
experiments.

The amplitude of inflationary gravitational waves is parametrized by
the tensor-to-scalar ratio, $r$, which is defined as the ratio of the
power in tensor modes to that in scalar modes at some pivot co-moving
scale, here taken to be $0.002$\,Mpc${}^{-1}$.  The combination of CMB
temperature data with measurements of large-scale structure sets the
current upper limit of $r < 0.22$ at $95\%$ confidence
\citep{2009ApJS..180..330K}.  However, cosmic variance ultimately
limits the $r$ constraints that can be obtained from total intensity
measurements, and further improvements will require direct measurement
of CMB polarization.

A detection of the dominant $E$-mode component of CMB polarization was first reported around
eight years ago \citep{DASI}, and since then many experiments have
reported further measurements \citep{CBIpol,2007ApJ...660..976S,
  CAPMAP,2008ApJ...684..771B, DASI3yr,BoomerangEE,
  2006ApJ...647..833P,2007ApJ...665...55W, 2007ApJS..170..335P}.  
  Recent observations have begun to characterize detailed features 
  in the $E$-mode spectrum
\citep{2009ApJ...705..978B, 2010ApJ...711.1123C, 2010arXiv1012.3191Q}. However, a 
measurement of the more interesting $B$-mode component will require 
a substantial improvement beyond present experimental sensitivities.

The weakness of the $B$-mode polarization signature presents a major 
experimental challenge. Aside from the question
of raw sensitivity, there are several other important obstacles that
must be overcome. Systematic errors must be controlled with
unprecedented accuracy if the small $B$-mode signal is not to be
degraded by confusion with the much larger $E$-mode and total 
intensity signals. Diffuse Galactic foregrounds will also
contaminate the observations, with synchrotron emission and thermal
dust emission expected to significantly complicate the interpretation
of polarization measurements across microwave frequencies.

\spider\ \citep{2008SPIE.7010E..79C, 2010SPIE.7741E..46F} is a
balloon-borne microwave polarimeter designed to search for the 
imprint of inflationary gravitational waves upon the polarization of the CMB.  From 
a vantage point above the bulk of the atmosphere, \spider\ will map
the CMB's polarization over a large portion of the sky with a resolution
close to one degree. \spider's main goal is to measure the amplitude of
gravitational waves with sensitivity down to $r = 0.03$ (with $99\%$
confidence), or to place an upper limit at this value if no detection is
made. To obtain this sensitivity, \spider\ will employ over 
$2500$ bolometers. 

\citet{2008ApJ...689..655M} investigated the potential impact of many
common sources of systematic error on the ability of \spider\ to achieve
its main scientific goal, setting tolerance limits to guide the design of the
instrument and observation strategy. In this paper, we continue this
work, moving our focus onto new systematic effects that have proven
to be important but that have so far received less attention. Instead
of setting tolerance limits, our aim is to ensure that the measured
performance of various instrument components and prototypes is
sufficient to meet our main goal.

We also develop a detailed model of the foreground emission \spider\
will see. This makes the sky model against which the systematic
effects are assessed more complete, and also enables us to further
test the robustness of our observation and analysis strategy. In this work we do
not attempt to address the important issue of separating the
foreground and CMB contributions to our data.
 
This paper is organized as follows. Section \ref{sec:instrument} gives
a brief overview of the instrument and introduces the systematic
effects we investigate. The model of polarized foregrounds we use is
described in Section~\ref{sec:skymodel}.  Sections~\ref{sec:simsmethod} 
and~\ref{sec:simsres} introduce the
simulation methodology and the baseline observation strategy we use, 
and present the results of these studies. Finally, Section~\ref{sec:conclusions} gives
a discussion of the results and our conclusions.

\section{The Instrument}
\label{sec:instrument}
A detailed description of the \spider\ instrument is presented in
\citet{2010SPIE.7741E..47R}. Here we provide a brief overview of the
instrument and the sources of error we consider.

The \spider\ payload consists of six separate monochromatic
instruments housed in a single liquid-helium cryostat. 
All six point in the
same direction on the sky.
The optical design is based on that of the successful \bicep\
telescope \citep{2010ApJ...711.1141T}. Each instrument includes a
telecentric refracting telescope consisting of two lenses cooled to $4$\,K. At
$150$\,GHz this design produces a beam on the sky with full-width at
half-maximum (\fwhm) close to $40$\,arcmin.
Each telescope focuses light onto a focal plane array of
antenna-coupled bolometers, which are cooled to $300$\,mK using
${}^3$He closed-cycle sorption fridges.
Each dual-polarization spatial pixel in the focal plane consists of
two phased arrays of slot dipole antennas, which are sensitive to
polarization along orthogonal axes. The power incident on these antennas
is deposited on to a bolometer and detected with a superconducting
transition-edge sensor (TES).  Superconducting quantum interference
devices (SQUIDS) are used to amplify the current produced by the TESs,
which is then read out using time-domain multiplexing \citep{2003RScI...74.3807D,2008JLTP..151..908B}. The focal-plane
architecture is described in more detail in
\citet{2008SPIE.7020E..38K}.  Note that the design of the
beam-defining optical components, the lenses and the focal plane
arrays, is very similar to that of the \biceptwo\ experiment
\citep{2010SPIE.7741E..40O}, although the architecture of the focal
plane unit is different; see Section~\ref{subsec:magnetic}.

A rotatable cold half-wave plate is mounted at the aperture of each
telescope, on the sky-side of the primary lens. The plates are
constructed from 330-mm diameter slabs of birefringent
single-crystal sapphire, with a quarter-wave quartz coating to minimize
reflections. The orientation of the half-wave plate determines the
orientation of the instrument's polarization sensitivity axes with respect to the sky (see equation
(\ref{eq:boloout})).  By making observations with the half-wave
plate oriented over a range of angles during the flight, high
polarization cross-linking can be achieved, improving the fidelity of
the recovered polarization maps.
The half-wave plates also provide powerful defense against several
important sources of systematic error.  Any error that does
not transform as a true polarization signal under rotation of the
half-wave plate can be distinguished from CMB signal during data analysis. 
This suppression will occur naturally using standard map-making
techniques (provided that the polarization cross-linking is sufficient), as
the systematic contributions to the data will have functional forms
nearly orthogonal to that of the true sky signal. If necessary,
further suppression may be achievable by modifying the map-making
algorithm to model the contribution to the data induced by the error,
and marginalizing over it if required.

In this paper, we consider several potential sources of systematic
error in the \spider\ instrument. First, the behavior of the half-wave plate
is weakly dependent on the frequency of the incident radiation, which
leads to non-idealities when averaged over the \spider\ bandwidths,
which are around $25\%$ of the central frequency.
%
Second, the focal plane is highly sensitive to external magnetic
fields, and the motion of the instrument through the magnetic field of
the Earth has the potential to generate significant spurious signals.
Using laboratory measurements of the magnetic pickup in the focal
plane, we estimate the in-flight response to the Earth's field.
%
Finally, we consider imperfections in the optical beams defined by the
antenna arrays and lenses, focusing on asymmetries and mismatches
between the orthogonally polarized detectors in each focal plane pixel.  
Optical characterization of the \spider\ and \biceptwo\ instruments
have shown that these beam systematics are present at a small level,
and we use these measurements to guide our studies of the expected
impact on \spider's science goals.

\section{Sky model: foregrounds}
\label{sec:skymodel}
Galactic foreground emission is expected to contribute significantly
to the polarized microwave emission across the sky, and may well
dominate the CMB gravitational-wave signal at all frequencies.
Therefore, it is important to include the foreground emission in our
sky model for this work. Furthermore, \spider\ will observe a large
fraction of the sky, potentially including regions close to the
Galactic plane where the foreground emission is very bright in
comparison to the CMB. The presence of this bright emission in our
data may affect the performance of the observation and analysis
strategy we use.

Unfortunately, these foregrounds are poorly constrained by current
data and poorly understood.  This is particularly true above around $90$\,GHz, where
the CMB emission is strongest and where \spider\ will operate. 
At these frequencies, the foreground emission is expected to be
dominated by thermal emission from interstellar dust.  Direct measurements by \Archeops~\citep{2004A&A...424..571B, 2005A&A...444..327P} and by \WMAP~\citep{2007ApJ...665..355K} have shown that this emission is polarized at the few percent level at frequencies higher than $90$~GHz.  This is in agreement with our theoretical understanding of the underlying physical mechanism~\citep{Draine:2009p8}, which can be summarized as follows.  Dust
grains are generally non-spherical, and preferentially emit radiation
polarized along their longest axis.  Mechanisms exist that align these
grains with this axis perpendicular to the Galactic magnetic field.  This leads to a net linear polarization of the observed emission.  Independent 
observations of the polarization of starlight~\citep{1996ApJ...462..316H, 2002ApJ...564..762F} are consistent with this picture.

Synchrotron emission, generated by the gyration of cosmic ray
electrons in the Galactic magnetic field, is intrinsically polarized
and constitutes the main polarized foreground at lower frequencies
\citep{2007ApJS..170..335P}.  Since the polarized emission from interstellar dust is expected to dominate the polarized Galactic synchrotron emission at frequencies above $\sim\!70$~GHz, we do not include the latter in our sky model.

Although little data is available regarding the polarized emission
from dust, the same is not true of its total intensity. In particular,
the IRAS satellite observed this emission across the sky at $100~\mu$m
and $240~\mu$m, close to the peak in the dust emission. By
constraining physically-motivated extrapolations of these observations
using further data, \citet[hereafter FDS]{1999ApJ...524..867F}
provided models of the emission at microwave wavelengths. At
$94$\,GHz, these models have been shown to agree well with the \WMAP~observations, up to a small normalization factor and some minor
structural differences in the Galactic plane
\citep{2009ApJS..180..265G}. Data are more limited in the higher frequency bands relevant to
\spider, but measurements agree well with the FDS predictions
\citep{2010ApJ...722.1057C, 2010ApJ...713..959V}. We use FDS's model 8
to trace the total
intensity of the dust~emission.

The other necessary fields for a full description of the dust emission, namely its degree and direction of polarization, are highly dependent on the
Galactic magnetic field. Since the observed polarization signal is the sum of the emission from
many independent regions along the line-of-sight, it is sensitive to
the three-dimensional structure of this magnetic field.
As a result, modeling the degree and direction of polarization of the dust emission requires the evaluation of appropriate line-of-sight integrals given three-dimensional models of the Galactic
magnetic field and of the other relevant Galactic constituents.

Away from the Galactic center, the Galactic magnetic field is usually
considered to have two nearly independent components: a large-scale
coherent field associated with the Galactic disk and a small-scale
field arising from turbulence in the interstellar plasma sourced by
astrophysical events such as supernovae and stellar winds. The most
informative probes of these fields are Faraday rotation measures of
pulsars and extra-Galactic radio sources
\citep{2006ApJS..167..230H,2006ApJ...642..868H}. Whilst there is
general agreement that the large-scale field follows a spiral pattern,
its detailed structure is still uncertain.
This uncertainty is unimportant when considering areas of sky at high 
Galactic latitudes: since the dust is concentrated in a thin disk
about the Galactic plane, we only see emission originating within around
$1$\,kpc or so of the Sun, a region in which the large-scale field is
reasonably well characterized. However, as \spider's large sky coverage may include part of the Galactic
plane, we require a model of the large-scale field structure in that
plane.
One popular candidate is the bisymmetric spiral
\citep{1994A&A...288..759H, 2008A&A...477..573S}, which can be written
as
\begin{align}
  B_{\rho} &= -B_0 \cos{\left(\phi + \psi
      \ln{\frac{\rho}{\rho_0}}\right)} \sin{p} \cos{\chi_0}  \nn\\
  B_{\phi} &= -B_0 \cos{\left(\phi + \psi
      \ln{\frac{\rho}{\rho_0}}\right)}  \cos{p} \cos{\chi_0} \nn\\
  B_{z} &= B_0 \sin{\chi_0}.
\label{eq:chapthree9}
\end{align}
Here, $\rho$, $\phi$ and $z$ are Galactocentric cylindrical
coordinates with $\phi$ measured from the direction of the Sun, $p$
is the pitch angle of the field, $\psi = 1/\tan{p}$, $\rho_0$ defines
the radial scale of the spiral, and $\chi_0$ parametrizes the
amplitude of the $z$ component. We use the parameter values suggested in
\citet{2008A&A...490.1093M}: $p = -8.5$\,degrees, $\rho_0 = 11$\,kpc
and $\chi_0 = 8$\,degrees, with the field amplitude set to $B_0 = 3~\mu$G and the distance between the Sun and the Galactic center
taken to be $8$\,kpc. Although this field model is unlikely to provide a full
description of our Galaxy \citep{2008A&A...486..819M,
  2008A&A...477..573S}, it is sufficient for our current purpose since we
do not require an accurate template of the sky, only a reasonable
approximation against which to test the performance of the experiment.

The turbulent component of the Galactic magnetic field is somewhat less well understood. When
constraining the above large-scale field, \citet{2008A&A...490.1093M}
simultaneously fit a small-scale field component with best-fit r.m.s\ amplitude
$B_{\mathrm{r.m.s.}}  = 1.7~\mu$G.  Since several different studies agree
that the r.m.s.\ amplitude is similar to the amplitude of the
large-scale field in the solar vicinity \citep{2002ApJ...564..762F,
  2006ApJ...642..868H}, we set $B_{\mathrm{r.m.s.}} = 2~\mu$G.  \citet{1996ApJ...458..194M} examined the rotation measures of
extra-Galactic sources across a small patch of sky and concluded that
the data were consistent with Kolmogorov turbulence on scales smaller
than $4$\,pc, assuming a statistically isotropic and homogeneous Gaussian
field. On larger scales, they found a somewhat flatter energy spectrum
with an outer scale of up to $96$\,pc. Kolmogorov-type spectra up to
kilo-parsec scales in the interstellar magnetic field and other
interstellar plasma components have also been reported by other
studies \citep{1995ApJ...443..209A, 2000ApJ...537..720L,
  2008arXiv0812.2023C}.
Since it is numerically intractable to generate a full-sky realization of this turbulent
field in three dimensions at sufficiently high resolution, we resort to independent
one-dimensional realizations along the line-of-sight to each
pixel. This model ignores correlations across the sky, but properly
incorporates the line-of-sight depolarization. We choose an injection
scale of $100$\,pc, assume the dissipation scale to be small, and use the
one-dimensional Kolmogorov energy spectral index of $-5/3$.

We model the large-scale spatial distribution of the dust density,
$n_d$, using a simplification of the model constrained in
\citet{2001ApJ...556..181D},
\begin{equation}
n_d = n_0 \exp{\left(-\frac{\rho}{\rho_d}\right)}
\sech^2{\left(\frac{z}{z_d}\right)}.
\label{eq:chapthree7}
\end{equation}
For consistency with the \WMAP~polarization analysis
\citep{2007ApJS..170..335P}, we take the scale height $z_d = 100$~pc
and the scale radius $\rho_d = 3$~kpc. We do not attempt to model the
small-scale variations in the dust density and temperature, which
may also affect the polarization degree and direction. Small-scale
variations in the total intensity are included via the FDS model.

The model also requires a description of the physics of grain
alignment and of the intrinsic polarization of the emission from an
individual grain. In general, these are complex functions of the
magnetic field and of various properties of the grains. Recently, good
progress has been made in describing the details of the alignment
using the theory of radiative torques \citep{2007MNRAS.378..910L,
  2008MNRAS.388..117H}. However, it is still difficult to produce a
well constrained quantitative description to apply to our model
\citep{2009arXiv0901.0146L}.
We instead describe the alignment in an integrated manner, without
recourse to the details of a particular physical mechanism. We assume
that the polarization direction is always perpendicular to the
component of the magnetic field in the plane of the sky, and that the
degree of polarization depends quadratically on the magnetic field
strength. This is similar to the behavior assumed in
\citet{2007ApJS..170..335P}.

We compute the Stokes parameter maps associated with our
three-dimensional model using the appropriate line-of-sight integrals,
\begin{align}
  I_{\thrD}(\theta, \phi) &=\epsilon(\nu) \int_0^{r_{\max}} n_d(\vect{r}) \, \ud r \nn\\
  Q_{\thrD}(\theta, \phi) &=\epsilon(\nu)\int_0^{r_{\max}} n_d(\vect{r}) p_0
  [B_\phi(\vect{r})^2 - B_\theta(\vect{r})^2] \, \ud r \nn\\
  U_{\thrD}(\theta, \phi) &= \epsilon(\nu)\int_0^{r_{\max}} n_d(\vect{r}) p_0
  [2B_\phi(\vect{r})B_\theta(\vect{r})]\, \ud r,
\label{}
\end{align}
where the normalization $p_0$ is chosen to reproduce the average
polarization fraction of $3.6\%$ reported by \WMAP~outside their P06 mask
\citep{2007ApJ...665..355K}. Here, $r$, $\theta$ and $\phi$ represent
spherical polar coordinates centered on the telescope, and $\epsilon$
is the emissivity of the dust as a function of frequency, $\nu$. Note
that we conform to the default convention applied in the
\healpix\footnote{See http://healpix.jpl.nasa.gov} package
\citep{2005ApJ...622..759G} regarding the sign of $U$.

From this model we require maps of the polarization direction,
$\gamma$, and degree, $P$, which are given by
\begin{align}
  P(\theta, \phi) &= \frac{\left( Q_{\thrD}^2 +
      U_{\thrD}^2\right)^{\frac{1}{2}}} { I_{\thrD}} \nn\\
  \gamma(\theta, \phi) &= \frac{1}{2}
  \arctan{\left(\frac{U_{\thrD}}{Q_{\thrD}}\right)}.
\end{align}
Finally, we scale by the FDS intensity maps to obtain our final dust model expressions:
\begin{align}
 I_{\dust} (\theta, \phi) &= I_{\fds}\nn\\
 Q_{\dust}(\theta, \phi) &= I_{\fds}P \cos{2 \gamma} \nn\\
 U_{\dust}(\theta, \phi) &= I_{\fds}P\sin{2 \gamma}.
\label{}
\end{align}

\begin{figure}
\begin{center}
  \includegraphics{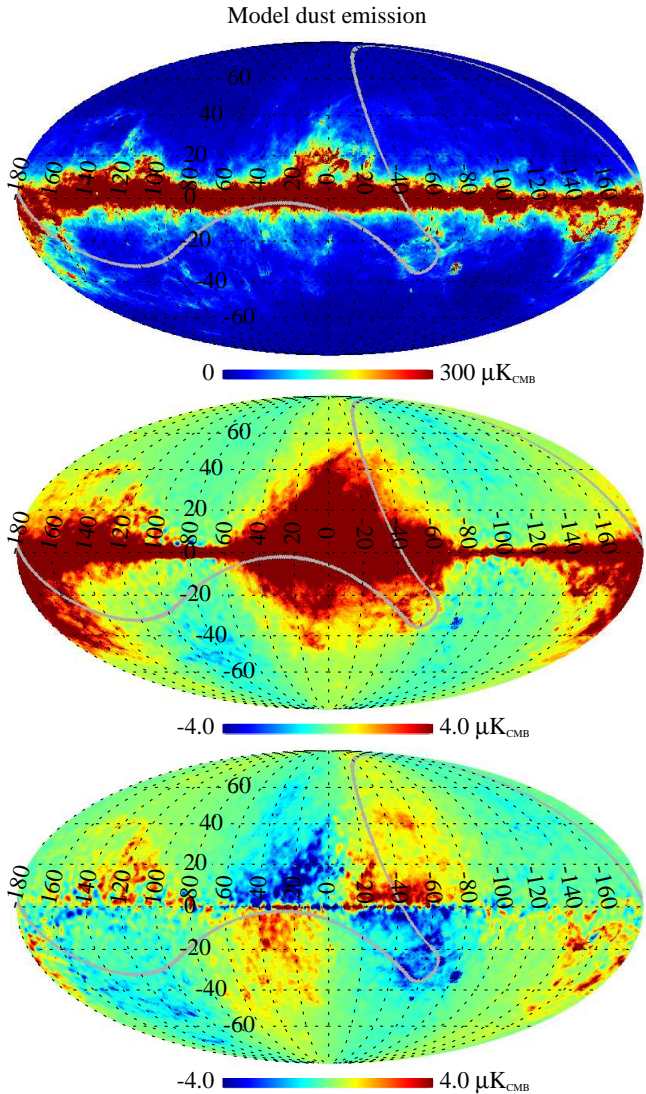}
  \caption[]{Stokes parameter maps (from top $I$, $Q$ and $U$) in Galactic
    coordinates for our
    model of thermal dust emission at $150$\,GHz. The grey lines show the region observed with the
    observation strategy described in Section \ref{sec:simsmethod},
    which excludes the Galactic center.}
  \label{fig:dustmaps}
\end{center}
\end{figure}

\begin{figure*}
\begin{center}
  \includegraphics{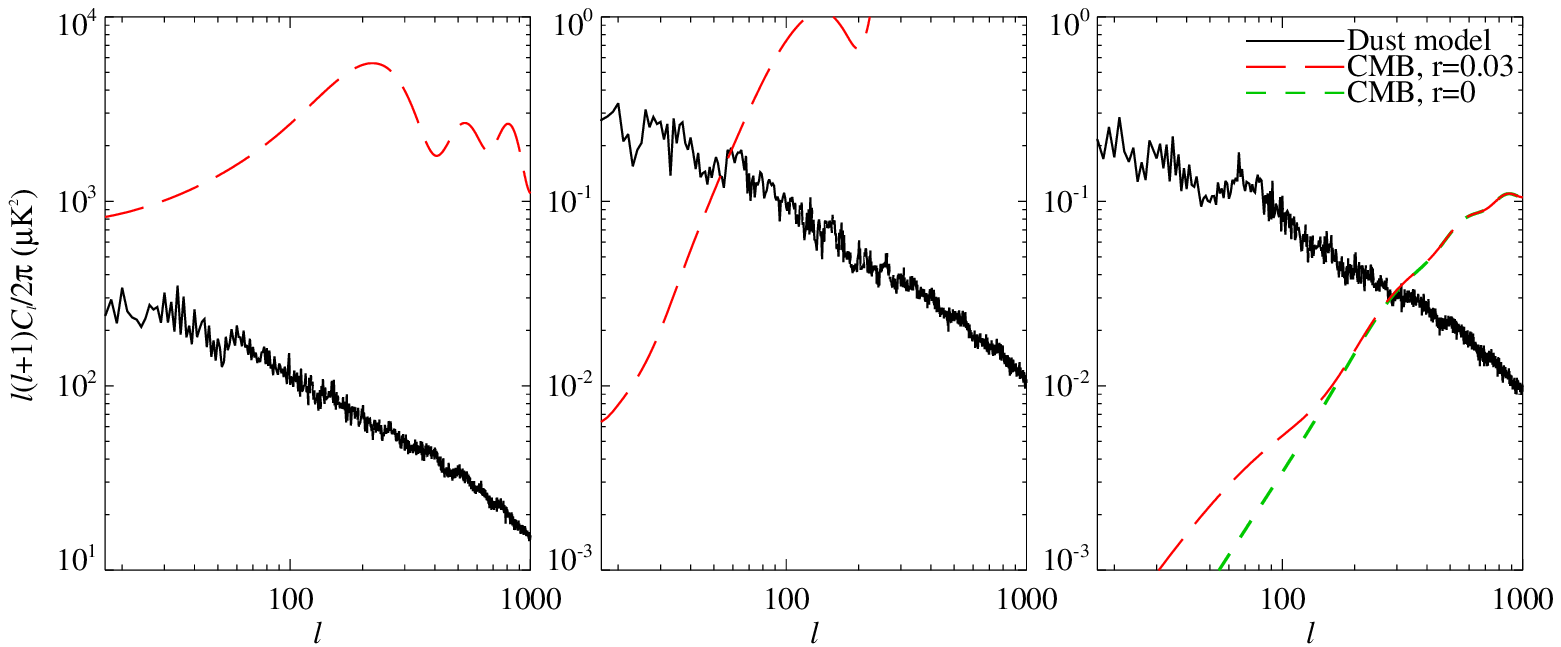}
  \caption[]{Power spectra (from left $I$, $E$ and $B$) of our model
    of thermal dust emission at $150$\,GHz. The CMB spectra for a
    universe consistent with current data with $r=0.03$ are shown for
    comparison, (red, long-dashed). In the right-hand panel, the
    contribution from gravitational lensing is also shown (green,
    dashed).}
  \label{fig:dustspectra}
\end{center}
\end{figure*}

Figure \ref{fig:dustmaps} shows the Stokes parameter fields for this
model evaluated at $150$\,GHz. In order to compare the dust emission to the
CMB, we plot the pseudo-power spectra of these maps (calculated using the
\healpix\ \anafast\ facility) in Figure \ref{fig:dustspectra}, masking regions within $10$~degrees of
the Galactic plane.

\section{Simulation method}
\label{sec:simsmethod}
We use the simulation pipeline and nominal observation and analysis
scheme introduced and investigated in \citet{2008ApJ...689..655M}.  This 
in turn is based largely on the analysis pipeline developed for
the $2003$ flight of \boomerang\ \citep{2006A&A...458..687M}, described
in \citet{2007A&A...470..771J}.

To limit the computational requirements, data are simulated for only
$8$ spatial focal-plane pixels ($16$ bolometers) in a single,
evenly-spaced column spanning one \spider\ focal plane. The two
antennas in each spatial pixel are sensitive to polarization parallel
and perpendicular to the column. This limited simulation is sufficient
for our current purpose, where we perform signal-only simulations and
investigate systematics effects which are unlikely to worsen as the
number of bolometers considered increases. Therefore, the
contamination generated by the various systematic errors reported here
can be treated as reasonable upper limits on the likely residuals in
the experiment itself.

We simulate a mid-November launch from Alice Springs, Australia (longitude
$128.5$\,degrees east, latitude $25.5$\,degrees south), with the
balloon subsequently drifting in longitude at $3.76 \times
10^{-4}$\,degrees per second.  The gondola is spun at
$36$\,degrees per second at a fixed elevation of $49$\,degrees, and we
observe only when the Sun is ten degrees or more below the
horizon. The half-wave plate is stepped once per day by
$22.5$\,degrees, and so we consider a four day observation period
which allows us to fully sample the polarization modulation the wave
plate induces; see equation (\ref{eq:boloout}). With this scanning
strategy, \spider\ observes approximately $60\%$ of the sky.

Based upon this scenario, the flight simulator generates the time-ordered
right ascension and declination observed by each spatial pixel, as well as
its orientation angle $\psi$ with respect to the sky basis. The sky model is
generated using the \healpix\ \synfast\ facility and the foreground
model described in Section \ref{sec:skymodel} using $n_\mathrm{side}
= 1024$, i.e.\ pixels of size $\sim 3.4$\,arcmin. The sky is also
smoothed with a Gaussian kernel with \fwhm\ of $40$\,arcmin to
account for the angular resolution of \spider. The model is generated
at $150$\,GHz, \spider's main science channel. As we are not concerned
with the detailed impact of instrumental noise or foreground
separation, this is sufficient for this work.
Note that we use the same CMB realization for each of the various simulations we perform.
The pointing information is then used to generate the time-ordered
data from the sky model and the wave-plate angle, $\theta$.  The
bolometer outputs are given by
\begin{align}
  d_\pm = & I_\sky \pm [Q_\sky \cos{2(\psi + 2\theta)} + U_\sky \sin{2(\psi +
    2\theta)} ],
\label{eq:boloout}
\end{align}
where the signs refer to the different bolometers in a spatial
pixel. Including systematic effects requires modifications to the
above procedures, which we describe in Section \ref{sec:simsres} for
the systematic errors we consider.

\spider\ maps are then made using an adaptation of the Jacobi iterative
map-maker as described in \citet{2007A&A...470..771J}. 
Although the simulations are noiseless, the maps are made using a
noise covariance matrix based on a $1/f$ spectrum with knee frequency
$100$\,mHz, a typical value for the \spider\ detectors
\citep{2010SPIE.7741E..47R}.
We also apply a one-pole high-pass filter at $10$\,mHz to the
time-ordered data during map-making, as this is expected to be
necessary for the real data to mitigate long-time scale systematic
errors such as incomplete knowledge of the system transfer functions.
Thus we properly include the expected loss of modes due to the low
frequency noise and filtering and the resulting degradation of the
recovered maps. These maps are produced at $n_\mathrm{side} = 256$,
i.e.\ a pixel size $\sim 13.7$\,arcmin.

By examining maps of the residuals, i.e.\ the difference between the
output \spider\ maps and the input model sky, we can begin to evaluate
the consequences of the various sources of error. In order to properly
assess their impact on our ability to meet our main goal of measuring
the gravitational-wave signal, however, it is necessary to proceed to the power
spectrum domain, where we can make direct comparisons with the
expected CMB spectra. 

We estimate spectra from our maps via a spherical harmonic transform
using \anafast. The decomposition of
polarization fields into $E$- and $B$-modes is not unique on the cut sky, however, and so
these so-called `pseudo'-spectra estimates contain mixing between the
two. The result of this is that the recovered $B$-mode spectrum is
biased by the much larger input $E$-mode spectrum
\citep{2003PhRvD..67b3501B, 2002PhRvD..65b3505L}.

If unaccounted for, this $E$--$B$ mixing dominates the systematic
contribution to the recovered $B$-mode spectrum. The development and
implementation of an optimal unbiased power spectrum estimator which
properly accounts for this mixing is beyond the scope of this
work. Instead, we eliminate this bias by estimating the pseudo-spectra
of the residual maps. This removes the true sky from our estimates,
allowing us to concentrate on the contribution from systematic
errors. We also lose a small contribution to the recovered $B$-mode
spectrum from map-level errors which correlate with the sky (e.g.\ an
absolute calibration error), but fully retain the much more worrisome
systematic effects which couple the recovered $B$-modes to $E$-modes
and total intensity. Using the power spectrum of the residuals has the
added advantage of eliminating the cosmic variance contribution from
the sky signal, allowing the systematic-induced bias to be accurately
determined without the need for many independent simulations.

The resulting power spectrum estimate,
$\widetilde{C}_l^{B\mathrm{,res}}$, should be compared to the expected
CMB $B$-mode pseudo-spectrum measured by \spider\ without foregrounds,
systematic errors or $E$--$B$ mixing. We estimate this by performing a
systematics- and dust-free simulation with no input $E$-modes and
calculating the resulting pseudo-spectrum, $\widetilde{C}_l^{B\mathrm{,noE}}$.
Finally, to aid interpretation we form $R_l$,
\begin{align}
R_l = \frac{\widetilde{C}_l^{B\mathrm{,res}}}{\widetilde{C}_l^{B\mathrm{,noE}}} C_l^B,
\end{align}
so that the final comparison can be made against the input CMB
spectrum, $C_l^B$. 

We emphasize that this procedure is not designed to give a complete
picture of \spider's ability to estimate the $B$-mode power spectrum
but to allow us to ensure that the systematic-induced bias is
small compared to the expected signal of interest.

\begin{figure}
\begin{center}
  \includegraphics{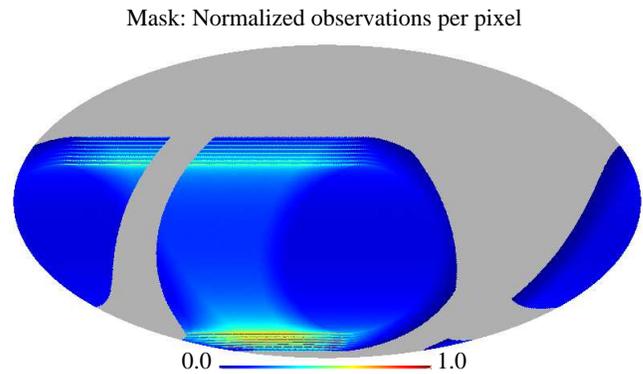}
  \caption[]{Map showing the sky-mask used when calculating power
    spectra in our simulation pipeline in equatorial coordinates.}
  \label{fig:mask}
\end{center}
\end{figure}

For the above power spectrum estimations we use a mask covering the
region observed, weighted by the time spent observing each sky-pixel
to reduce the impact of poorly sampled pixels near the map edges. We
also add a $\pm 10 $\,degree Galactic-plane mask, as this area is heavily obscured by foregrounds and will not
be available for cosmological analysis. This mask is shown in Figure \ref{fig:mask}.

\section{Simulation results}
\label{sec:simsres}
\subsection{Dust model}
We begin by investigating whether the addition of dust to our sky
model reduces the effectiveness of the nominal observation and
analysis strategy described in Section \ref{sec:simsmethod}. To do so
we perform two simulations, one using our full sky model and one with
a CMB-only sky. Figure \ref{fig:rl_dust} compares the residual
spectra, $R_l$, from these simulations. As demonstrated in
\citet{2008ApJ...689..655M}, ten map-making iterations is sufficient
to reduce the residuals to a negligible level for the CMB-only
case. Using our updated sky model, the residuals remain significant after ten map-making iterations, with $R_l$ greater than the CMB
$B$-mode spectrum for $r=0.03$ on large scales ($l < 20$).  We
find that increasing the number of iterations to around $30$
suppresses the residuals to a negligible level. Figure
\ref{fig:qres_dust} shows the residual $Q$ map for this case (the
residual $U$ map has similar amplitude), and demonstrates that the
recovered maps are of high quality, with little loss of information
arising from the time-stream filtering and $1/f$ noise. Note that the
small Galactic regions with relatively high residuals are behind our
Galactic mask. In accordance with these results, we use $30$ map-making
iterations for the remaining simulations reported in this paper,
unless otherwise stated.
For completeness, Figure \ref{fig:rl_dust} also shows $R_l$ after $60$
iterations, at which point the map-maker has fully converged.
\beginfigure
\begin{center}
  \includegraphics{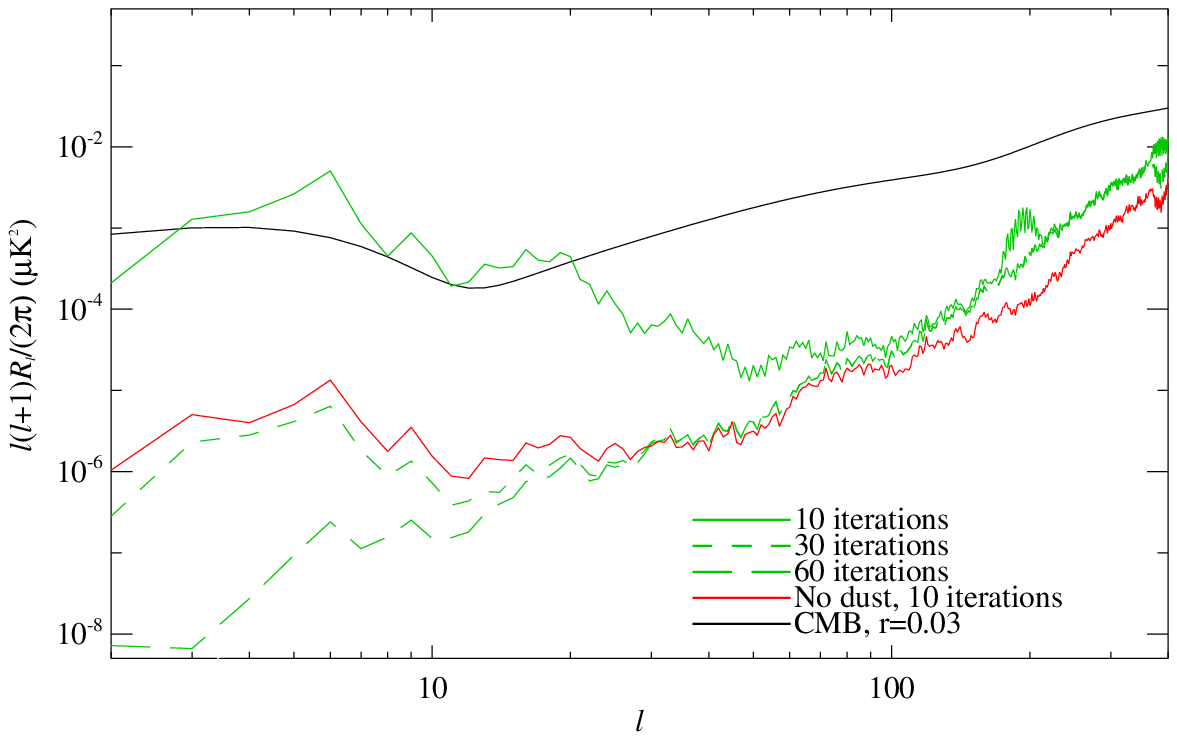}
  \caption[]{Residual $B$-mode spectra, $R_l$, for simulations run
    with and without the dust emission (green and red respectively)
    after ten iterations of the map-maker.  Residuals for simulations
    with dust emission after $30$ (dashed) and $60$ (long-dashed)
    iterations are also shown. The CMB $B$-mode spectrum for $r=0.03$
    is shown for comparison (black).  This confirms that our previous
    nominal observation strategy is not compromised by the inclusion
    of the dust emission, with residuals orders-of-magnitude below the
    level of the signal of interest after $30$ iterations.}
  \label{fig:rl_dust}
\end{center}
\efigure
\beginfigure
\begin{center}
  \includegraphics{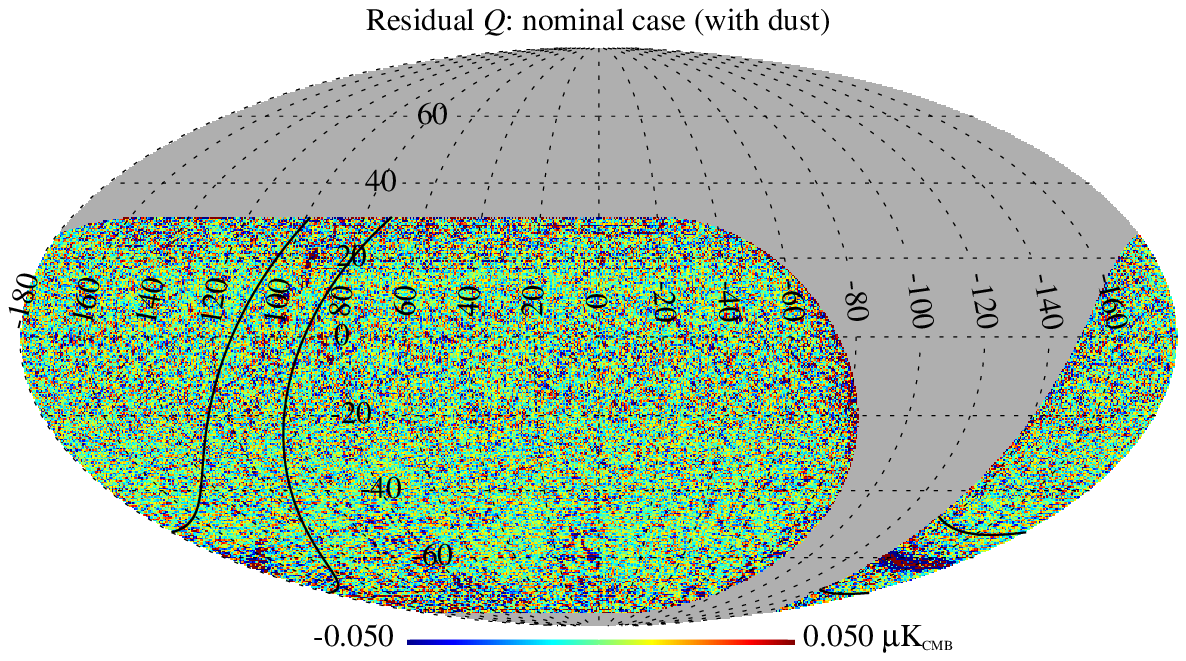}
  \caption[]{Residual $Q$ polarization map for a simulation run with
    our complete sky model (CMB and dust) after $30$ iterations of the
    map-maker in equatorial coordinates. The black lines show the
    Galactic cut used when calculating power spectra. The residuals
    are low across the observed sky apart from some small regions in
    the Galactic plane.}
  \label{fig:qres_dust}
\end{center}
\efigure

\subsection{Non-ideal half-wave plate}
With the sky model thus established, we turn to our investigation of 
the impact of instrumental systematics on \spider's expected sensitivity.  We begin 
with the non-trivial spectral response of
the \spider\ half-wave plate.

The \spider\ half-wave plates consist of birefringent sapphire with
thickness chosen such that the optical path length difference along
the crystal axes is exactly a half wave-length at the center of the
appropriate frequency band. The anti-reflection quartz coating is also
optimized to this frequency.
However, as we move away from this central frequency, the plate
behavior becomes non-ideal; the optical path difference is no longer
exactly a half wave-length, and the amplitude of reflections, although
still suppressed, increases.  When averaged over the frequency band
these non-idealities remain.
Furthermore, the band-averaged non-idealities depend on the emission
spectrum, $S(\nu)$, of the incident radiation.  This spectrum is \textit{a
  priori} unknown, as we do not know the relative contributions from
the CMB and the various foregrounds. This complicates efforts to
predict the effects of the half-wave plate non-idealities, and to
account for them during data analysis should it prove necessary.

We use the analytic formula derived in \citet{2010ApOpt..49.6313B} to
calculate the band-averaged \Muller matrix describing the \spider\
half-wave plate for both CMB and dust emission spectra. This formula
is based on a physical optics model of the half-wave plate.  We use
the refractive index measurements reported in
\citet{2010SPIE.7741E..64B} for sapphire at $5$K, and take the
quartz anti-reflection coatings to have a refractive index of
$1.95$. In performing the band-averaging, the dust emission spectrum
is modeled following model number eight in FDS, using the average
temperatures for the two model dust components, $T_1 = 9.7K$ and $T_2
= 16.2K$, and so
\begin{align}
  S_{\mathrm{CMB}}(\nu) =& \left. \frac{\ud B(T, \nu)}{\ud T}
  \right|_{T_\mathrm{CMB}} \nn\\
  S_{\mathrm{dust}}(\nu) =&
  \left(\frac{\nu}{\nu_0}\right)^{1.67}B(T_1, \nu) + 0.933
  \left(\frac{\nu}{\nu_0} \right)^{2.7} B(T_2,\nu),
\end{align}
where $B(T, \nu)$ is the black-body spectrum and $\nu_0$ is
$150$\,GHz. We also incorporate the \spider\ detector response spectrum
reported in \citet{2010SPIE.7741E..46F}.
The band-averaged half-wave plate \Muller matrices are fully described
by four independent parameters such that
  \begin{align}
    \mM_{\mathrm{HWP,band}} =
    \left(\begin{array}{cccc} 
        T & p & 0 & 0  \\
        p & T & 0 & 0  \\
        0 & 0 & c & -s \\
        0 & 0 & s & c
\end{array} \right).
\label{eq:hwpparams}
\end{align}
Some of these parameters have straight forward interpretations. $T$
represents an overall transmission loss, and $p$ arises due to
differences between the transmission spectra for radiation polarized
along the two crystal axes. The calculated parameter values for our
$150$\,GHz band are given in Table \ref{table:table1}, although note
that as the CMB is expected to have no circular polarization, the $s$
parameter should not be relevant to \spider.
\begin{table}
\begin{center}
\begin{tabular}{ccccc}
\hline
         & $T$            & $p$            & $c$               & $s$            \\
\hline
Ideal &            $1$ &             $0$ &            $-1$ &              $0$  \\    
CMB  & $0.97127$ & $0.00673$ & $-0.94157$  &   $0.03126$  \\    
Dust & $0.96952$ & $0.00628$ & $-0.94067$  & $-0.01190$  \\    
\hline
\end{tabular}
\end{center}
\caption[]{
  \rm Parameters describing the band-averaged half-wave plate for a \spider\
  detector response function, and the expected emission spectra for the CMB
  anisotropies and thermal dust. Values are also given for an ideal half-wave plate.}
\label{table:table1}
\end{table}


The \Muller matrix describing the transmission of radiation through
the entire instrument from the sky to a bolometer can be expressed in
terms of $\mM_{\mathrm{HWP,band}} $ and the matrix describing
the antenna polarization, $\mM_{\mathrm{ant}}$, as
\begin{align}
  \mM_{\mathrm{ins}} =&\mM_{\mathrm{ant}} \mR_{\epsilon}
  \mR_{-\theta} \mM_{\mathrm{HWP,band}} \mR_{\theta}
  \mR_{\psi},
\end{align}
where $\epsilon$ is the angle between the antenna's polarization
direction and $\psi$ i.e.\ its orientation in the focal plane (here
taken to be $0$ and $90$\,degrees for the two bolometers in a spatial
pixel), and the matrix $\mR_{\alpha}$ describes a basis rotation
through an angle $\alpha$.  We assume the polarization of the antenna
is perfect and so
\begin{align}
  \mM_{\mathrm{ant}} = \frac{1}{2}\left(\begin{array}{cccc}
      1 & 1 & 0 & 0  \\
      1 & 1 & 0 & 0  \\
      0 & 0 & 0 & 0 \\
      0 & 0 & 0 & 0
\end{array} \right).
\end{align}
The bolometers measure the total power incident upon them, and so the
ideal bolometer outputs, equation (\ref{eq:boloout}), are replaced by
\citep{2010ApOpt..49.6313B}
%
%
\begin{align}
  4d_\pm & = \, 2 T I_\sky  \nn\\
  \pm & (T-c)[Q_\sky \cos{2(\psi + 2\theta)}  + U_\sky \sin{2(\psi + 2\theta)} ] \nn\\
   \pm & 2p[I_\sky \cos{2\theta} - Q_\sky \cos{2(\psi + \theta)} - U_\sky \sin{2(\psi + \theta)} ] \nn\\
   \pm & (T+c)[Q_\sky \cos{2\psi} + U_\sky \sin{2\psi}].
  \label{eq:boloouthwp}
\end{align}
To add the half-wave plate non-idealities to our simulations we make
this replacement when generating the time-ordered data.

Examining equation (\ref{eq:boloouthwp}), we see that the total
intensity contribution is reduced by the transmission loss caused by
the half-wave plate, as expected. Similarly, the polarization
efficiency is reduced to $(T-c)/2$. Both of these effects are
degenerate with the more general calibration requirements of the
experiment. Of greater interest are the remaining contributions to the
bolometer outputs, which have dependencies on the half-wave
plate orientation unlike those of the ideal outputs, being either independent of
$\theta$ or sinusoidal in $2\theta$.
As these new contributions transform differently under rotation of the
half-wave plate to the ideal bolometer outputs, these errors should be
suppressed at the map level by the half-wave plate modulation. However,
our nominal four day scan is not sufficient to fully sample these
terms to take full advantage of this suppression, and so for this
section we increase our scan duration to eight days.

In the following simulation results, we calibrate the \spider\ maps
produced by the pipeline by applying the appropriate factor [for
polarization $2/(T-c)$] before calculating the residual maps and
spectra. This enables us to investigate the more complicated and less
well understood systematics introduced by the half-wave plate. Although
the calibration factor varies between the sky components, the difference
is small ($0.1\%$) in comparison to our absolute calibration
requirements and to the likely level of foreground residuals after
foreground separation.

As the half-wave plate parameters vary between the sky components, we
perform two independent simulations for the CMB and dust. The residual
spectra for these simulations are shown in Figure
\ref{fig:rl_hwp}. For the dust-only simulation the residuals are not
negligible, rising to over $40\%$ of the expected CMB $B$-mode
spectrum for $r=0.03$ at low $l$. The CMB-only residual is lower, but
rises to around $20\%$ of the expected CMB spectrum.  The residual $Q$
map for the dust-only simulation is shown in Figure \ref{fig:qres_hwp}
(the residual $U$ map has similar amplitude).
By running a further dust-only simulation on an unpolarized sky
(i.e.\ with $Q = U= 0$ everywhere) and setting $T$ and $c$ to their
ideal values, we see that the residuals are dominated by leakage of
the dust total intensity into polarization via $p$, as shown in
Figure \ref{fig:rl_hwp}. Similar simulations isolating the contribution
from the $(T+c)/2$ terms in equation (\ref{eq:boloouthwp}) confirm
that such an error does not lead to significant residuals for any
reasonable parameter values. Note that after $30$ iterations, the
map-maker has fully converged for these simulations.
\beginfigure
\begin{center}
  \includegraphics{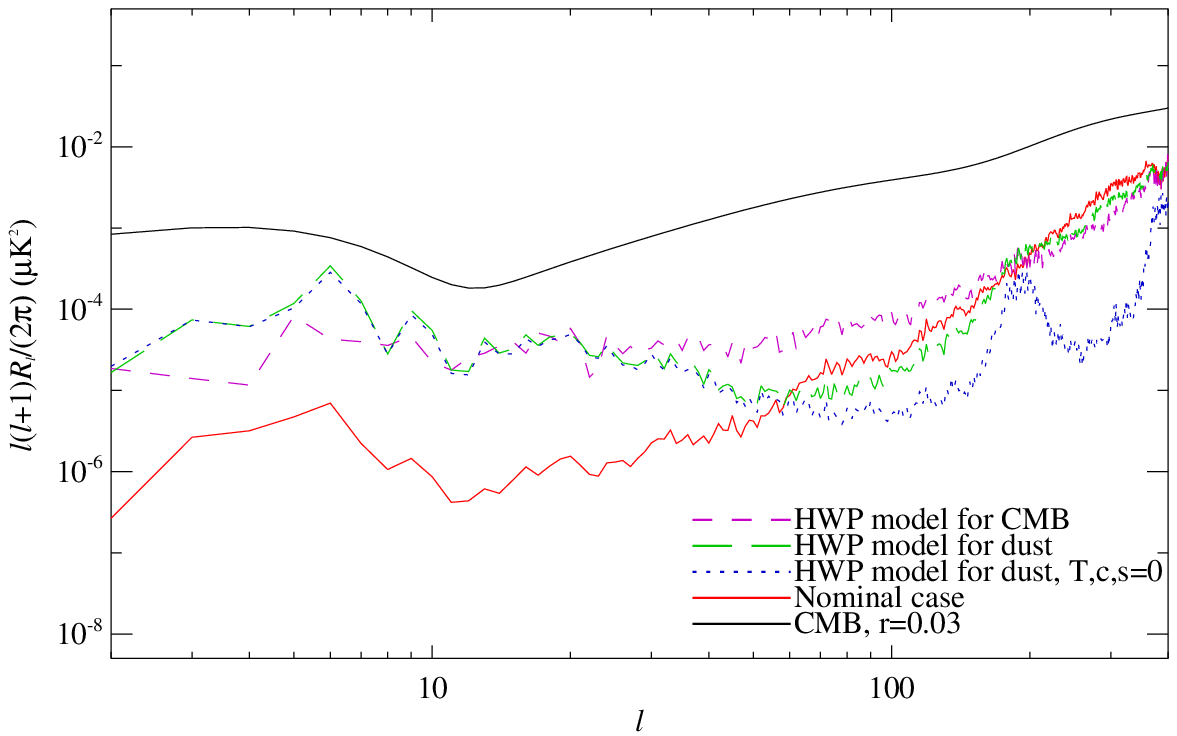}
  \caption[]{Residual $B$-mode spectra, $R_l$, for simulations using
    the non-ideal half-wave plate model for the CMB (magenta, dashed)
    and dust (green, long-dashed). The spectrum for our nominal
    simulation (red) and the CMB $B$-mode spectrum for $r=0.03$
    (black) are shown for comparison. The spectrum for a simulation
    with $p = p_{\mathrm{dust}}$ and only the dust total intensity on
    the sky (i.e.\ no polarization) is also shown (dotted blue).}
  \label{fig:rl_hwp}
\end{center}
\efigure
\beginfigure
\begin{center}
  \includegraphics{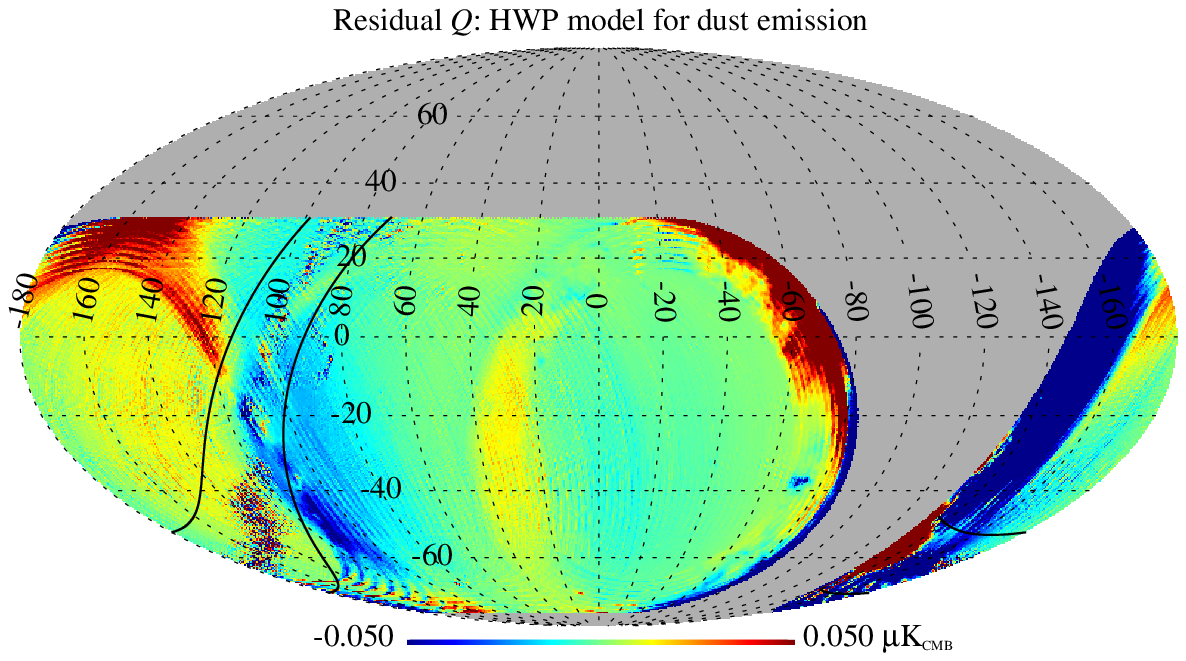}
  \caption[]{Residual $Q$ map for a dust-only simulation including the
    half-wave plate model in equatorial coordinates. The black lines
    show the Galactic cut used when calculating power spectra.}
\label{fig:qres_hwp}
\end{center}
\efigure

Note that taking the unrealistic step of masking the Galactic plane on the
\emph{input} sky significantly reduces the residuals resulting from
the half-wave plate non-idealities.  Although the plane is always
masked in the \spider\ maps when we calculate spectra, the filtering
of the time-ordered data partially delocalizes the large systematic
signal generated as the instrument scans across the plane, moving the
resulting residuals to high latitude regions of the
map.

Our simulations have shown that the errors introduced by our non-ideal
half-wave plate model have the potential to mildly compromise
\spider's science results if not accounted for during map-making.
However, the relevant parameters describing the half-wave plate
non-idealities are not a strong function of the emission spectrum of
the sky: comparing the values for the CMB and average dust spectra,
the most troublesome parameter, $p$, changes by only around ten
percent, and $T$ and $c$ change by less than one percent. Similarly,
variation in these parameters across the sky (as, for example, the
dust temperatures vary) are expected to be smaller still.
Therefore, we expect to be able to accurately correct for these errors
during map-making by upgrading the algorithm to include the extra
contributions to the bolometer outputs in the pointing matrix, using
constant calibrated values of $T$, $p$ and $c$, as proposed in
\citet{2010ApOpt..49.6313B}. This should reduce the map-level
residuals by at least an order of magnitude, i.e.\ to a negligible
level given our current science goals. For future experiments
targeting $r< 0.01$, more detailed simulations will need to be
undertaken to investigate the viability of this strategy.
Modifying the map-maker in this manner may also provide greater
flexibility in \spider's scanning strategy, as the increase from a
$4$-day cycle in our wave plate angle to an $8$-day cycle might not be
required.

\subsection{Magnetic field sensitivity}
\label{subsec:magnetic}
A second potential concern for \spider\ is stray magnetic field pickup in the
superconducting electronics chain.  The SQUID amplifiers used in the signal readout chain are
essentially highly sensitive magnetometers; left unshielded,
they will not only amplify signal from the bolometers but also from the
Earth's magnetic field.  Furthermore, the transition temperature, $T_c$, of
the TES detectors is weakly dependent on the magnetic field environment.  Motion of the gondola through the terrestrial magnetic field may thus
induce spurious signals in the \spider\ time streams. In
order to mitigate these effects, considerable effort has been made to
shield the focal plane so that pickup from the Earth's field will be
sub-dominant to the CMB.  \citet{2010SPIE.7741E..47R} describes recent
changes to the \spider\ design, notably a move from a flat focal plane
architecture to a shielded box scheme, with the aim of significantly
reducing the magnetic pickup.

These improvements have been successful, and the magnetic response of 
a \spider\ telescope is now low enough
that it has proven difficult to characterize through laboratory
measurements. Response amplitudes have been measured for many of the detectors
for the three magnetic field axes, with the remaining amplitudes being
too small to detect against the measurement noise.  These
measurements were taken with the TES bolometers on transition (as they would be in flight), so that any changes in either
$T_c$ or SQUID flux would be detected.  In order to assign amplitudes to the
eight detector pairs in our simulation, we have selected randomly from
among the detected response amplitudes, taking care to ensure that the
variation in amplitudes across the focal plane and across the
different magnetic field axes is typical of that seen in the
measurements.

It is not only the overall amplitude of the pickup that is important;
any difference in magnetic response within a detector pair will lead to spurious
polarization signals. We have therefore preserved detector pairings
when selecting the amplitudes for our model. However, as these differences are
typically much smaller than the overall amplitude, they are not well
constrained by our measurements. In reality the differences may be
significantly smaller, and so the simulations presented here represent
an upper limit on the spurious signal \spider\ is likely to see.

Due to the very low signal levels, characterizing the frequency
dependence of the pickup is not feasible, so instead we use a spectrum
typical of those measured for the previous focal plane design: a
one-pole filter with a time constant of $0.3$\,s.

To include magnetic pickup in our simulations, the response amplitudes and modeled
frequency dependence are used to convert the input Earth's magnetic
field (in Tesla) to the output stray pickup observed in the \spider\
time streams (in $\mu$K).  The Earth's magnetic field is determined for
each \spider\ pointing (latitude, longitude, altitude and time) using
the {\it World Magnetic Model} code and data provided by the {\it
  National Geophysical Data
  Center}\footnote{http://www.ngdc.noaa.gov/geomag/WMM/}, which
estimates the strength and direction of the Earth's main magnetic
field for a given point. We use the pointing time streams to account for the 
changing orientation of the focal plane with respect to the terrestrial field.  
In order to isolate the systematic error
induced by the pickup, for these simulations the only input signal is
the Earth's magnetic field; the CMB and dust sky signals are not
included. These magnetic field time-ordered data are the inputs to the
map-maker. In this case there is no differencing of the final map with
an input map; the output stray field map is treated as the residual
map.

Figure \ref{fig:qres_mag} shows the resulting map for $Q$
polarization signal induced by the Earth's magnetic field.  The $U$
map has similar amplitude.  Figure \ref{fig:rl_mag} illustrates this
magnetic response in multipole space. The power spectrum residuals,
$R_l$, are less than $3\%$ of the CMB $B$-mode spectrum for $r=0.03$
across all multipoles, and so we conclude that with the improved
focal-plane shielding, the magnetic pickup will not affect \spider's
science goals.
\beginfigure
\begin{center}
   \includegraphics{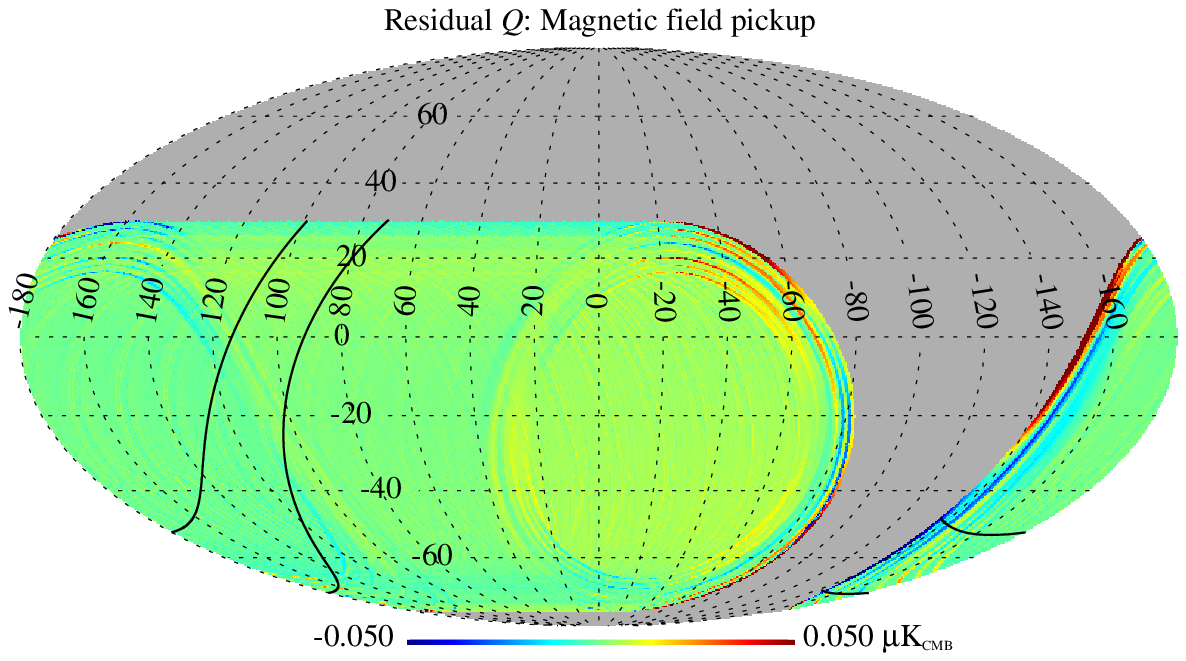}
   \caption[]{Residual $Q$ map for a simulation including the signal
     induced by the Earth's magnetic field in equatorial
     coordinates. The black lines show the Galactic cut used
       when calculating power spectra.  The magnetic field for each
     \spider\ pointing was determined using the {\it World Magnetic
       Model} code and data provided by the {\it National Geophysical
       Data Center}.  The resulting time-ordered magnetic field data
     were used to construct the observed $Q$ map.}
\label{fig:qres_mag}
\end{center}
\efigure
\beginfigure
\begin{center}
   \includegraphics[]{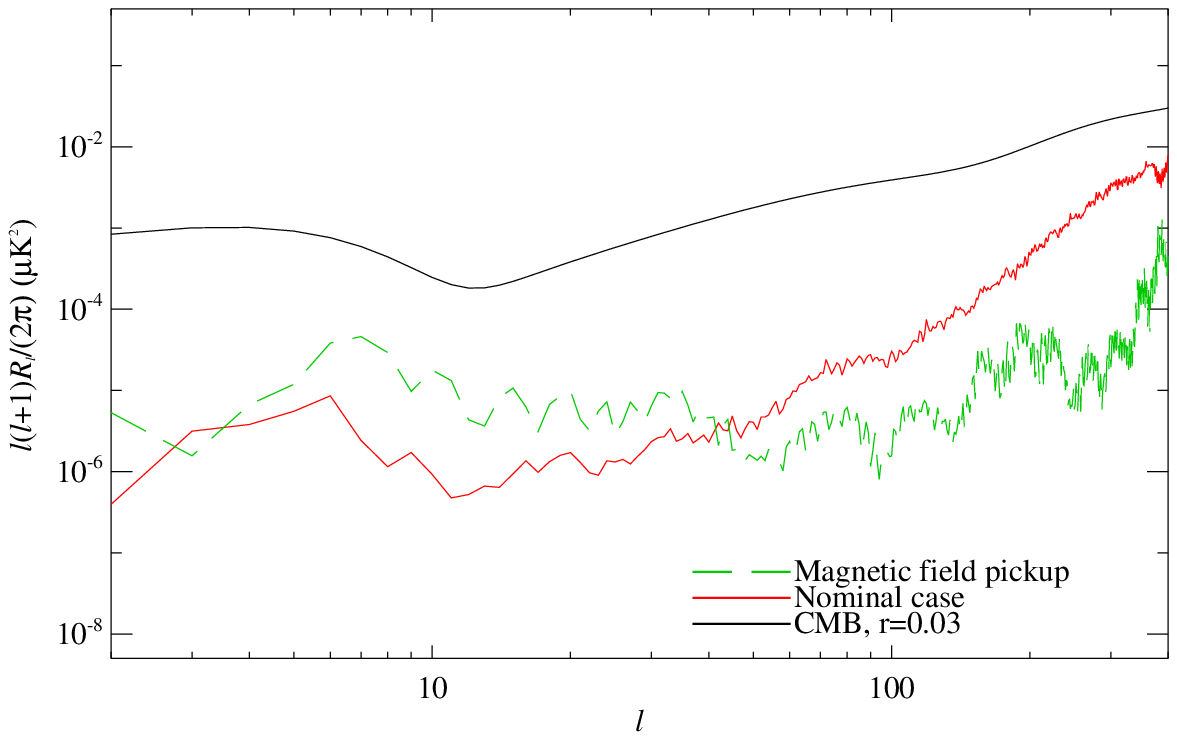}
   \caption[]{Residual spectrum, $R_l$, for a simulation of the signal
     induced by the Earth's magnetic field (green, long-dashed). The
     spectrum for our nominal simulation (red) and the CMB $B$-mode
     spectrum for $r=0.03$ (black) are shown for comparison.}
   \label{fig:rl_mag}
\end{center}
\efigure

\subsection{Mis-matched beams}
In this section we examine the impact of non-idealities in the beams
produced by the \spider\ optical system. Ideally, the instrumental
beams would be azimuthally symmetric and would be independent of the
antenna polarization.
We consider several departures from this ideal situation,
motivated by reported characterizations of the \biceptwo\ instrument
(which shares a common optical and antenna design with \spider) and
early measurements of the \spider\ beams. In particular we use the
results presented in \citet{2010SPIE.7741E..23A} to guide us. Of most
concern are mismatches between the beams seen by the two bolometers
in a spatial pixel, as these have the potential to strongly couple
total intensity on the sky into the observed polarization. In this work we
consider three such effects that have proven to describe accurately
the main departures from ideality in the measured beams: a
differential pointing error of $1.5$\,arcmin, a differential
ellipticity of $0.012$ and a differential beam width of $0.5\%$.
The method we employ to include these effects in our
simulations is designed to be sufficiently flexible to be applied to a
more detailed description of the beams, such as will be available when
full beam measurements of the \spider\ instruments are carried out in the
near future.

Our baseline simulation method assumes that the instrumental beams are
ideal, i.e.\ that each bolometer couples to the sky via the same azimuthally
symmetric Gaussian kernel. This azimuthal symmetry ensures that the
computationally-intensive convolution operation need only be performed
once for each sky pixel. For a general kernel this is not the case:
the convolution is a function of the instrument orientation, which is
constantly changing throughout the scan. Performing a convolution for
each sample in the time-ordered data is not possible in a reasonable
time. Therefore, to investigate the impact of non-azimuthally
symmetric beams, we model them as the sum of a small number of
Gaussians, each offset relative to the nominal pointing direction and
with differing widths and amplitudes,
\begin{align}
B(\vx) = \sum_{n=1}^N a_n G(\vx; \vx_n, \sigma_n).
\label{eq:beammodel}
\end{align}
Here $B(\vx)$ is the beam as a function of position on the sky when
the instrument is in some fiducial orientation, $G(\vx; \vx_0,
\sigma)$ is a unit-normalized Gaussian centered on $\vx_0$ with width
$\sigma$, and the amplitudes, $a_n$, sum to unity. As well as tracking
the location of the nominal pointing direction on the sky, the
pointing generator has been modified to also track the location of the
center of each sub-beam, so that the full beam can be reconstructed at
all times in the appropriate location and orientation.

To generate the bolometer outputs for this model, the contribution
from each (azimuthally symmetric) sub-beam, $d_n$, is first calculated
using equation (\ref{eq:boloout}) or equation (\ref{eq:boloouthwp}),
along with the sky-model convolved with a Gaussian of width
$\sigma_n$. As the convolution operator is linear, the final
bolometer output is then given by
\begin{align}
d = \sum_{n=1}^N a_n d_n.
\end{align}
This method only requires a factor of $N$ more convolution operations
(or less if some of the $\sigma_n$ are equal to each other), as well
as a factor of $N$ more operations in the pointing and time-ordered
data generation.
Note that, in keeping with the general approach of this paper, no
attempt is made to correct for these beam errors during the map-making
and power spectrum analysis.

So far, we have described a method to include non-azimuthally
symmetric beams in our simulations. We also wish to consider the
impact of differences in the beams seen by the two bolometers in each
spatial pixel. This can be achieved by using different parameter values
($N$, $\sigma_n$ and $\vx_n$) for the two beams, $B_1(\vx)$ and
$B_2(\vx)$.  We can
also allow the beams to vary across the focal plane by using different
beam models for different focal-plane locations.

We must be careful with the applicability of this model. It is
important that we only consider two different antenna polarization
orientations across our focal plane. For other polarization
orientations (e.g. at $\pm 45$\,degrees), the instrument beams are
functions of $B_1(\vx)$ and $B_2(\vx)$ in a manner which does not
generally conform to equation (\ref{eq:beammodel}).  Similarly, this
model does not include (and cannot in general be extended to include)
cross-couplings between the two polarization states induced by the
optical system.

We perform simulations for each of the three beam mismatch errors we
consider. In each case, we model the beams to have errors typical of
those reported in \citet{2010SPIE.7741E..23A}. This represents the
worst case for the likely performance of the \spider\ instruments, as
further research and development is expected to improve the fidelity
of the beams produced. Note that, in this section only, we slightly
reduce the ideal beam \fwhm\ used in the simulations to match those
measured for \biceptwo, from $40$ to $31$\,arcmin.

We begin by considering differential beam pointing, i.e.\ differences
in the beam centers seen by the two bolometers in a spatial
pixel. Such errors couple gradients in the total intensity of the
radiation incident on the optical system into linear polarization in
the focal plane. If such contamination is not removed, the related
tolerance limits are strict \citep{2003PhRvD..67d3004H,
  2007MNRAS.376.1767O}.  \spider\ was designed with such
concerns in mind, however: the half-wave plate is on the sky side of all the
optical components. The effect of this coupling on the
bolometer outputs will thus be independent of the half-wave plate
orientation, and so will be suppressed in the map domain to a degree
dependent on the half-wave plate rotation strategy.
Note that unlike the pointing errors considered in
\citet{2008ApJ...689..655M} where the pointing properties were defined
to be time-invariant on the sky, here they are defined to be
time-invariant in the focal plane.

Modeling the differential pointing is straightforward: we simply use
one `sub-beam' for each bolometer in a spatial pixel, with an
appropriate offset between them.  For this offset we use a typical value of
$1.5$\,arcmin. The residual spectrum for a simulation using this
model is shown in Figure \ref{fig:rl_beams}, and the residual $Q$ map
is shown in Figure \ref{fig:qres_point} (the residual $U$ map has
similar amplitude). On large scales, where the gravitational wave
signal resides, the residuals are small, rising to around $20\%$ of
the CMB $B$-mode spectrum for $r=0.03$ at $l \sim 100$.  On smaller
scales the residuals are significant, becoming comparable to the CMB
spectrum around $l\sim 300$.  This does not compromise
\spider's goals, however, as the gravitational lensing signal and instrumental noise are
expected to dominate on these scales.
Note that the residuals do not depend strongly on the orientation of
the pointing error in the focal plane. Here we present the case where
the pointing errors are parallel to the column of detectors we
simulate.  Repeating the simulation with the pointing errors
perpendicular to the column, the residuals are generally lower for $l
< 150$, and similar on smaller scales.

Comparing these low residuals to the strict tolerance limits
found for differential pointing errors in the absence of polarization
modulation with a half-wave plate \citep{2003PhRvD..67d3004H,
  2007MNRAS.376.1767O}, we can conclude that our half-wave plate
rotation strategy leads to strong suppression of these errors at the
map-making stage.

Next we consider differential beam ellipticity, which also couples the
total intensity of the radiation incident on the optical system into
linear polarization in the focal plane, through a local quadrupolar
pattern.  Defining the ellipticity of a beam as $e =
(\fwhm_{\mathrm{maj}} - \fwhm_{\mathrm{min}}) / (\fwhm_{\mathrm{maj}}
+ \fwhm_{\mathrm{min}})$, where the subscripts denote the major and
minor axes of the ellipse, we take a typical value of the difference
in ellipticity between the beams for the two bolometers in a spatial
pixel of $e_1 - e_2 = 0.012$.

We model an elliptical beam using two sub-beams with identical widths,
$\sigma = 31$\,arcmin, and centers displaced from the nominal
pointing direction by $\delta/2$ in opposite directions along the
desired orientation of the major axis. We find that for $\delta =
2$\,arcmin, $e = 0.006$. Figure \ref{fig:ellipbeams} shows the
difference between this model and a true elliptical Gaussian with
matching ellipticity; note that they agree to $0.1\%$ everywhere. To
produce the desired differential ellipticities in our simulations, we
use this model with the major axis parallel to the polarization
direction for each antenna.
\beginfigure
\begin{center}
  \includegraphics{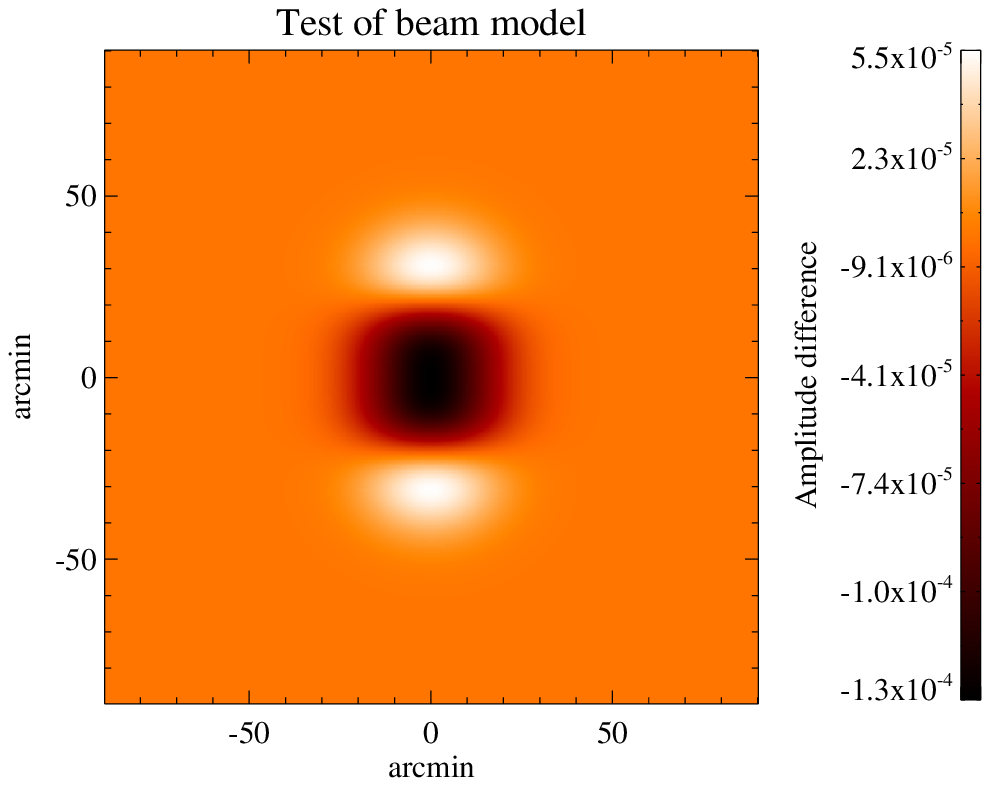}
  \caption[]{Map showing the difference between the elliptical-beam
    model we implement in our simulations, with ellipticity $e=0.006$,
    and a true elliptical Gaussian. The beams are normalized such that
    the true elliptical Gaussian peaks at one. Note that the
    difference is small, peaking at around $0.1\%$. }
  \label{fig:ellipbeams}
\end{center}
\efigure
The resulting residuals, shown in Figure \ref{fig:rl_beams}, 
are negligible across the multipole range relevant to
\spider. 
\beginfigure
\begin{center}
  \includegraphics{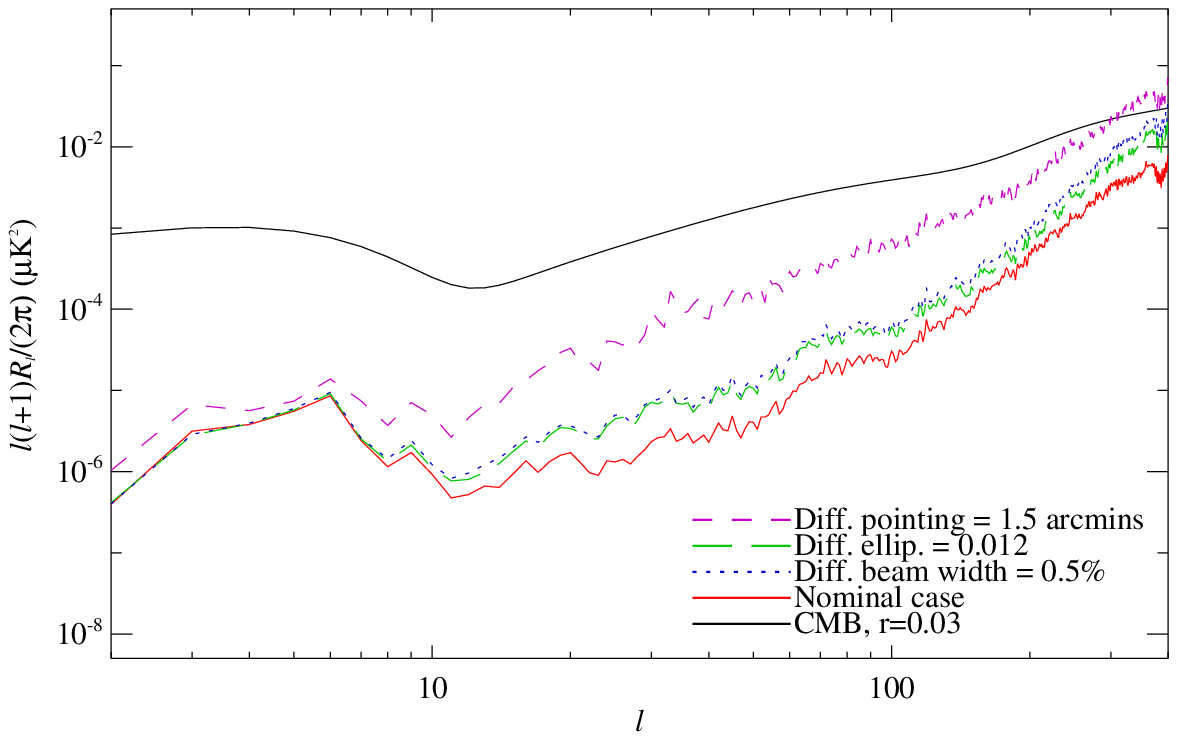}
  \caption[]{Residual spectra, $R_l$, for simulations with a
    differential pointing error of $1.5$\,arcmin (magenta, dashed), a
    differential ellipticity error of $0.012$ (green, long-dashed) and
    a differential beam width of $0.5\%$ (blue, dotted). Note
    that the orientation of the pointing errors and ellipticities in
    the focal plane do not significantly affect the residuals. The
    spectrum for our nominal simulation (red), and the CMB $B$-mode
    spectrum for $r=0.03$ (black) are shown for comparison.}
  \label{fig:rl_beams}
\end{center}
\efigure
\beginfigure
\begin{center}
  \includegraphics{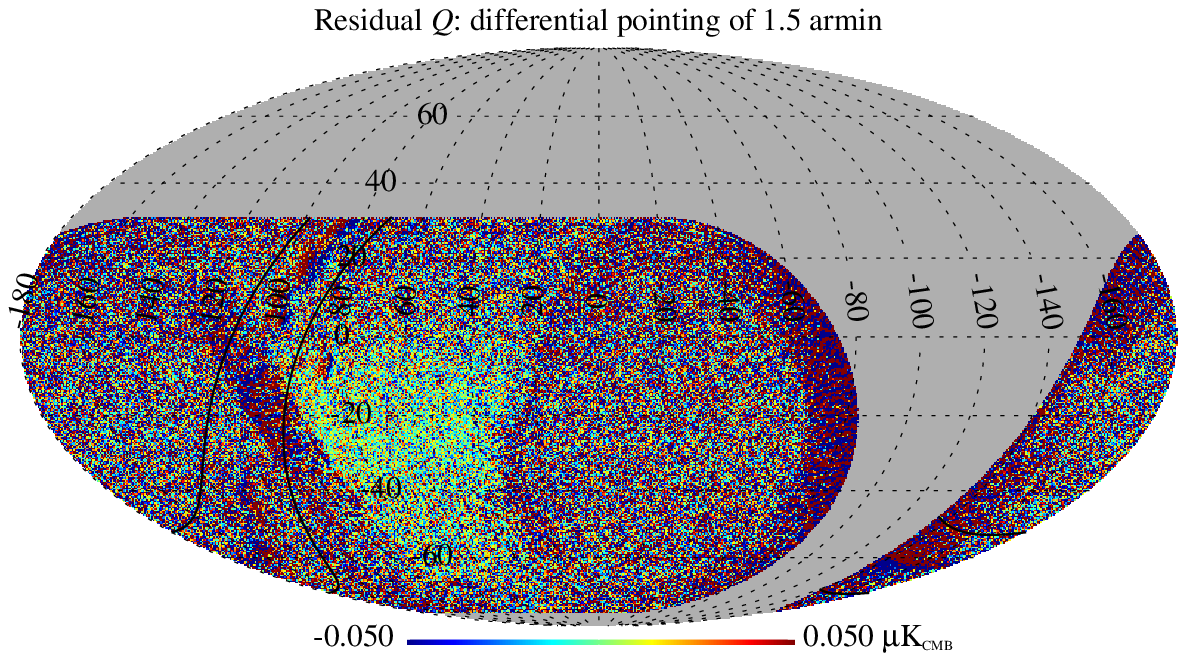}
  \caption[]{Residual $Q$ map for a simulation with a differential
    pointing error of $1.5$\,arcmin in equatorial coordinates. The
    black lines show the Galactic cut used when calculating power
    spectra..}
\label{fig:qres_point}
\end{center}
\efigure

The relative orientation of the major axes of the elliptical beams and
the polarization of the antennas is also an important factor in
determining the impact of a differential ellipticity error. In the
absence of half-wave plate modulation, differential ellipticity
contributes a signal indistinguishable from a true polarization
signal, i.e.\ one which cannot be mitigated through instrument
rotation. The nature of this signal depends on the relative
orientation of the major axes and the antenna polarization: if the
major axes of the beams are at $45$\,degrees to the corresponding
antenna-polarization orientations this signal produces a $B$-mode
pattern in the recovered maps \citep{2008PhRvD..77h3003S}. Therefore,
we also perform a simulation with the major axes of the elliptical
beams at $\pm 45$\,degrees to the antenna polarization
orientations. This simulation produces residuals comparable to those
shown in Figure \ref{fig:rl_beams}, confirming that our half-wave plate
rotation strategy is sufficient to mitigate the expected differential
ellipticity in the optical system.

Finally, we consider differential beam widths, i.e.\ differences in
the widths of the beams for the bolometers in a spatial pixel, taking
a typical value from \citet{2010SPIE.7741E..23A} of $0.5\%$. To
do so, we simply use one `sub-beam' with the appropriate beam-width
for each bolometer. We find that the residuals are negligible across
all multipoles, with $R_l$ remaining within $20\%$ of its value
in our baseline simulation; see Figure \ref{fig:rl_beams}.

We have shown that the expected beam mismatch errors in the \spider\
optical system will not compromise the science goals of the
experiment, even if no attempt is made to correct for them during data
analysis, as the half-wave plate modulation works well in mitigating
them at the map-making stage. Although we have not considered
variations in the magnitude and orientation of these mismatches
across the focal plane, such variation is only likely to reduce the
overall residuals, as the contributions from different spatial pixels
combine with less coherence.

\section{Conclusions}
\label{sec:conclusions}
We have considered several potential sources of systematic error in
\spider\ and used simulations to assess their likely impact on the
experiment's main science goal, assuming that no attempt is made to correct 
the effects during data analysis. Unlike previous examinations of
systematic errors, the goal of this work is not to set tolerance limits
to guide the design of the instrument, but instead to use measurements
of particular instrumental non-idealities to assess their
importance. Many of the systematics considered here have so far
received little attention in the literature.
We have also described a model of the polarized Galactic foreground
emission \spider\ will observe, based on a three-dimensional model of
the Galactic magnetic field and dust, and we have demonstrated that
its presence will not compromise the performance of our baseline
observation strategy.

We have considered three main sources of systematic errors that have
been characterized through measurements of \spider\ (or \spider-like)
hardware; the non-ideal spectral response of the half-wave plate,
spurious signals generated by the motion of the instrument through the
Earth's magnetic field, and mismatches in the instrumental beams
generated by orthogonally polarized antennas.

We concluded that non-idealities in the spectral response of the
half-wave plates result in errors in the recovered $B$-mode power
spectrum that may mildly bias our measurement of $r$
if the gravitational wave amplitude is close to our target sensitivity
of $r=0.03$ or below.  This systematic effect, left uncorrected, will not significantly
degrade the scientific results of the experiment, nor lead to a
spurious detection of gravitational waves.  However, it should be
straightforward to remove this bias by introducing three parameters
describing the half-wave plate non-idealities (which can be accurately
measured pre-flight) into the pointing matrix during map-making.

Our simulations of the stray magnetic field pickup in the TES detectors 
and SQUID amplifiers have 
demonstrated that the recent changes to the focal plane design to
improve the magnetic shielding \citep{2010SPIE.7741E..47R} are
sufficient to suppress the pickup to a negligible level.  Using 
laboratory measurements of the focal plane magnetic field response,
including differential pickup within detector pairs, the simulated
$B$-mode power spectrum residual is over an order of magnitude smaller
than the CMB for $r=0.03$ across the multipoles of interest.

Finally, we have shown that the differential beam non-idealities
measured in \spider\ and \biceptwo, which has a similar optical 
system, are not significant; the modulation of these errors introduced
by the half-wave plate on the sky side of all the optical components
ensures that our science goals are not affected by their presence. The
method introduced here will also be used to perform similar tests on
more sophisticated descriptions of the optical response of the
\spider\ instruments when they become available.

We have considered these sources of error in their application to
\spider, but the simulation methodology we have developed here is also
applicable to other experiments with similar instrument components
attempting to measure the imprint of gravitational waves on the
polarization of the CMB.

\acknowledgments 
The author acknowledges support from STFC under the standard grant
scheme (PP/E002129). The \spider~project is supported by NASA award
NNX07AL64G. WCJ acknowledges the support of the Alfred P. Sloan
Foundation. Some of the results in this paper have been derived using
the \healpix\ package~\citep{2005ApJ...622..759G} as well as the FFTW
package~\citep{1386650}.

\bibliography{references}
\end{document}